\documentclass[twocolumn, prl, superscriptaddress]{revtex4-1}
\usepackage{amssymb} 
\usepackage{amsmath}
\usepackage{epsfig}
\usepackage{graphics, graphicx}
\usepackage{bbold}
\usepackage{psfrag}
\usepackage{mathcomp}
\usepackage{subfigure}
\usepackage{verbatim}
\usepackage{color}
\usepackage[colorlinks,citecolor=blue]{hyperref}

\begin{document}

\title{Three in Many: \ Efimov physics in the presence of a Fermi sea}
\author{Mingyuan Sun}
\email{msun@connect.ust.hk}
\affiliation{Institute for Advanced Study, Tsinghua University, Beijing, 100084, China}
\author{Xiaoling Cui}
\email{xlcui@iphy.ac.cn}
\affiliation{Beijing National Laboratory for Condensed Matter Physics, Institute of Physics, Chinese Academy of Sciences, Beijing, 100190, China}
\affiliation{Songshan Lake Materials Laboratory , Dongguan, Guangdong 523808, China}
\date{\today}

\begin{abstract}
Motivated by recent experiments on $^{6}$Li-$^{133}$Cs atomic mixtures with high mass imbalance, we study the Efimov correlation in atomic system of two heavy bosons ($^{133}$Cs) immersed in  a bath of light fermions ($^{6}$Li). Using the Born-Oppenheimer approximation, we identify two different regimes, depending on the Fermi momentum of light fermions ($k_F$) and the boson-fermion scattering length $a_s(<0)$, where the presence of underlying Fermi sea plays distinct roles in 
the Efimov-type binding of bosons.   Namely, in the regime $k_F|a_s|\lesssim1$ ($k_F|a_s|\gtrsim1$), 
the Fermi sea induces an attractive (repulsive) effective interaction between bosons and thus favors (disfavors) the formation of bound state, which can be seen as the Efimov trimer dressed by the fermion cloud. Interestingly, this implies a non-monotonic behavior of these bound states as increasing the fermion density (or $k_F$). 
Moreover, we establish a generalized universal scaling law for the emergence/variation of such dressed Efimov bound states when incorporating a new scale ($k_F$) brought by the Fermi sea. These results can be directly  tested in Li-Cs cold atoms experiment  by measuring the modified bound state spectrum and the shifted Efimov resonance, which manifest an emergent non-trivial Efimov correlation in a fermionic many-body environment.   
\end{abstract}
\maketitle

A main task of physics research is uncovering novel few-body correlations in interacting many-body systems. In 1970, Vitaly Efimov discovers an intriguing  three-body effect, in which three identical bosons can form a sequence of bound states in the vicinity of two-body resonance and they satisfy a universal scaling law \cite{Efimov}.  In recent years, the Efimov effect has been generalized to a variety of three-body systems with different statistics and mass ratios, and has also been verified in the ultracold gases of various atomic mixtures\cite{Braaten,Greene,Naidon1, Efimov_Exp0,Efimov_Exp1,Efimov_Exp1bu,Efimov_Exp3,Efimov_Exp4, Efimov_Exp5,Efimov_Exp9,Efimov_Exp10,Efimov_Exp6,Efimov_Exp7, Efimov_Exp8,Efimov_Exp11,rf_1,rf_2,scaling_1,scaling_2,scaling_3, Helium}. Given these developments, it is time to move on to address the interplay of such novel three-body effect and a many-body environment. For instance, how to visualize the Efimov effect/correlation in a many-body system? and in turn, how would a many-body background change the Efimov physics? 
These are all very charming while highly challenging problems, because they are associated with non-trivial few-body correlations in the strongly interacting many-body systems.

The polaron system, which consists of a few impurities immersed in a majority of other particles, can serve as the simplest platform for investigating the interplay problem between  few and many. 
Depending on the statistics of majority particles, the system can be classified into Bose polarons and Fermi polarons, both having been realized in cold atoms experiments\cite{Zwierlein,Salomon,Grimm,Kohl,Grimm2016,Roati,Arhus,JILA,Lamb}. Theoretically, there have been a number of  studies revealing the universal three-body correlations in Bose polaron and Fermi polaron systems \cite{Levinsen1, Levinsen2, Cui3, Cui4, Levinsen3, Naidon,Parish, Zhou, Zinner1, Endo, Nishida, Cui1, Cui2, Zinner3, Zinner4}. 
Nevertheless, the experimental detection of three-body correlation in polaron systems still remains at its infancy\cite{footnote}. 

In this work, we reveal the Efimov correlation in the impurity system consisting of two heavy bosons and a bath of light fermions. Our study is motivated by recent experimental progresses on $^{6}$Li(fermion)-$^{133}$Cs(boson) atomic mixtures\cite{scaling_2,scaling_3,LiCs1,LiCs2,LiCs3,LiCs4,LiCs5,LiCs6,Chin}. Such system has a number of unique properties. 
First, the high mass ratio between bosons and fermions ($133/6$) supports a small Efimov scaling factor in Cs-Cs-Li three-body system, which enables the resolution of three consecutive Efimov resonances in realistic experiments\cite{scaling_2,scaling_3}. Secondly, by tuning the relative number of Li and Cs, one can prepare either the Bose polaron or the Fermi polaron in such mixture system. In our previous work, we have revealed visible Efimov signature in the spectral response of Bose polarons consisting of one Li impurity and a bath of Cs atoms\cite{Cui3,Cui4}. In the present work, we will study an alternative case, i.e., the Fermi polaron system consisting of two Cs atoms and a bath of Li fermions, and address how the Efimov-type bound state is modified by the presence of a Fermi sea background. Previous studies of similar setup showed that the Efimov bound state will be disfavored by the  Fermi sea due to the Pauli blocking effect\cite{Zhou, Zinner1}  and the screening effect\cite{Endo}.

\begin{figure}[t]
\includegraphics[width=8.0cm,height=13.0cm]{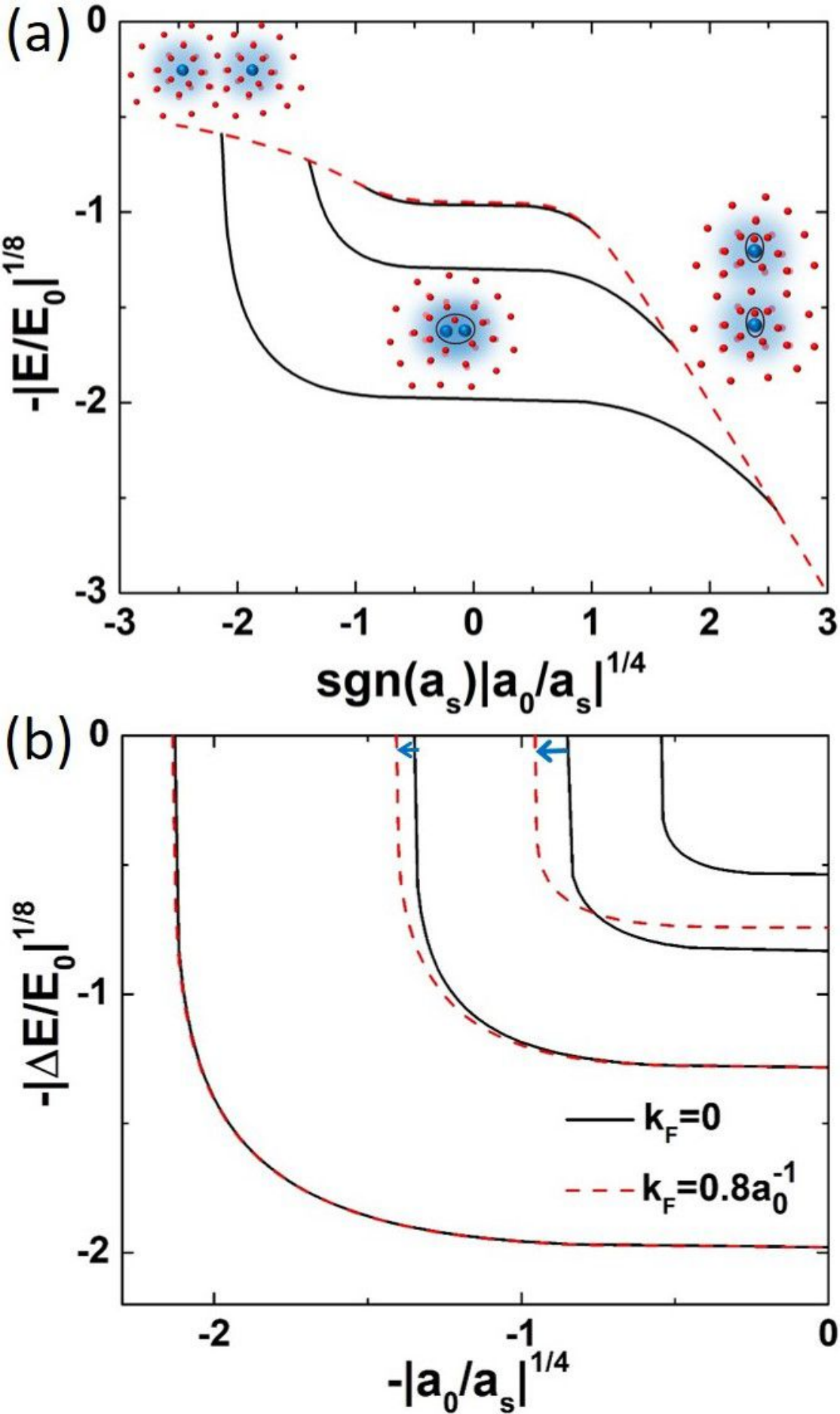}
\caption{(Color Online).  (a) Energy spectrum of two $^{133}$Cs bosons immersed in a bath of $^6$Li fermions with Fermi momentum $k_F=0.8a_0^{-1}$. Dashed line shows twice of single polaron energy (i.e., $2E_{p}$). 
(b) Bound state spectrum $\Delta E\equiv E-2E_{p}$ with a finite $k_F=0.8 a_0^{-1}$ (red dashed lines), in comparison to the Efimov spectrum (black solid lines) of three-body system (i.e., $k_F=0$).  
}  \label{fig1}
\end{figure}

Here we study the dressed bound state between two Cs bosons and the rest Li fermions using the Born-Oppenheimer approximation(BOA). The resulted typical energy spectrum is shown in Fig.\ref{fig1}. 
One can see that the presence of a Fermi sea can significantly modify the bound state spectrum and the discrete Efimov scaling. We identify two regimes, depending on the dimensionless parameter $k_F|a_s|$  ($k_F$ is the Fermi momentum and $a_s(<0)$ is the boson-fermion scattering length), where the presence of Fermi sea plays distinct roles in the Efimov-type binding of bosons.  Namely, 
in the regime $k_F|a_s|\lesssim1$ the Fermi sea will favor the bound state formation due to the enhanced boson-boson attraction mediated by fermions; while in the regime $k_F|a_s|\gtrsim1$, the Pauli-blocking effect dominates and the Fermi sea disfavors the bound state formation. Such opposite effects of fermion background immediately suggest a non-monotonic behavior of the bound state property (including the locations of their appearance and their binding energies) as increasing the fermion density, which can be directly probed  in experiments.  
Finally, we establish a generalized universal scaling law for the emergence/variation of these dressed  bound states by incorporating the Fermi sea effect. These results review an emergent Efimov correlation in a fermionic many-body system, which represents an important interplay effect between few-body physics and many-body environment.
 

{\it Model.} We consider a system of two heavy bosons (with mass $M$) interacting with a bath of light fermions (with mass $m\ (\ll M)$), which is described by the Hamiltonian ($\hbar=1$ throughout the paper)
\begin{equation}
  \mathcal{H}=\frac{{\bf P}_{1}^2}{2M}+\frac{{\bf P}_{2}^2}{2M}+\sum\limits_{j=1}^{N}\frac{{\bf p}_j^2}{2m}+\sum_{j=1}^{N}\Big( V({\bf R}_{1}-{\bf r}_j)+V({\bf R}_{2}-{\bf r}_j) \Big),
\end{equation}
where ${\bf R}_{1(2)}$ and ${\bf P}_{1(2)}$ label the position and momentum of the two  boson atoms, while ${\bf r}_j$ and ${\bf p}_j$ ($j=1,\dots,N$) label the position and momentum of $N$ majority fermions; $V({\bf r})$ is the boson-fermion interaction potential described by an $s$-wave scattering length $a_s$, which can be tuned across resonance. Here we have neglected the weak boson-boson background interaction for simplicity. 

Given the large mass ratio between Cs and Li, $M/m=133/6$, we adopt the Born-Oppenheimer approximation(BOA). The validity of BOA can be tested by comparing the Efimov scaling factor of Li-Cs-Cs system ($5.28$ under BOA) with the accurate value $4.87$, which only gives a small relative error $\sim8\%$. Previously, the BOA has been used to obtain 
the effective interaction between two heavy particles that is mediated by fermions\cite{Nishida1}. Namely, given two particles with fixed coordinates ${\bf R}_1$ and ${\bf R}_2$, the interaction energy  of the system is determined by the spectral shift of light fermions from both the bound states and the scattering states:  
\begin{equation}
  E(R)=-\frac{\kappa_+^2+\kappa_-^2}{2m}-\int_0^{k_F}dkk\frac{\delta_+(k)+\delta_-(k)}{m\pi}
  \label{energy}
\end{equation}
here $R=|{\bf R}_1-{\bf R}_2|$ is the relative distance, and the bound state solution $\kappa_{\pm}$ and the scattering phase shifts $\delta_{\pm}$ can all be obtained by matching the according wave functions to Bethe-Pierls boundary condition at $|r-R_{1,2}|\rightarrow 0$. By subtracting the asymptotic limit for two uncorrelated impurities, where $E(R\rightarrow\infty)$ approaches twice the single polaron energy $2E_{p}$, one can define the effective interaction between two impurities as 
\begin{equation}
  V_{\rm eff}(R)=E(R)-2E_{p}. \label{potential}
\end{equation}


Given the effective potential in Eq.~\ref{potential}, we complete the last step of solving the bound states of the two bosons from the Schr{\" o}dinger equation:
\begin{equation}
\left( \frac{P^2}{M} + V_{\rm eff}(R) \right) \Psi({\bf R}) = \Delta E \Psi({\bf R}).
 \label{SchrodingerEq}
\end{equation}
Here $\Delta E$ is the binding energy of two bosons and the total interaction energy of the system is given by $E=\Delta E+2E_p$. In numerics, we have set a short-range cutoff at $R=R_c$ with $\Psi(R_c)=0$. Throughout the paper we choose the length unit as $a_0=200R_c$ and accordingly the energy unit as $E_0=1/(ma_0^2)$. 

The typical eigen-spectrum is shown in Fig.\ref{fig1} for the system of two Cs impurities imbedded in a bath of Li fermions with $k_F=0.8a_0^{-1}$.   From  Fig.\ref{fig1}(a),   we can see that in the weak coupling regime $E$ follows $2E_p$, indicating two independent polarons. As increasing $1/a_s$ beyond critical values $1/a_c^{(n)}$ $(n=1,2...)$, a sequence of discrete levels emerge below the two-polaron continuum, i.e., $E<2E_{p}$, representing the formation of dressed bound states involving two bosons and the surrounding Fermi cloud. In the deep molecule limit, these bound states merge into the dimer-dimer continuum with energy $\sim 2\epsilon_b= -1/(\mu a_s^2)$ (here $\mu=mM/(m+M)$ is the reduced mass). In this limit, each boson is tightly bounded with essentially one fermion to form a molecule and two molecules are nearly uncorrelated.

In Fig.\ref{fig1}(b), we further plot the bound state spectrum $\Delta E$ in $1/a_s<0$ side for a typical $k_F$, in comparison with the Efimov spectrum of three-body system (corresponding to $k_F=0$). One can see that the presence of Fermi sea with a finite $k_F$  can impose distinct effects to the shallow bound states depending on the interaction regime. For bound states far from resonance (i.e., small $|a_s|$), the presence of Fermi sea can facilitate the bound state formation, in that $|\Delta E|$ gets larger at a given $a_s$ and the critical $a_c^{(n)}$ moves to smaller value (see left arrows in Fig.\ref{fig1}(b)), i.e., weaker coupling strength\cite{footnote1}. However, for the bound states near resonance (large $|a_s|$), they will be destroyed by the Fermi sea such that no corresponding discrete levels emerge below the two-polaron continuum. 

\begin{figure}[t]
\includegraphics[width=8.0cm,height=13.0cm]{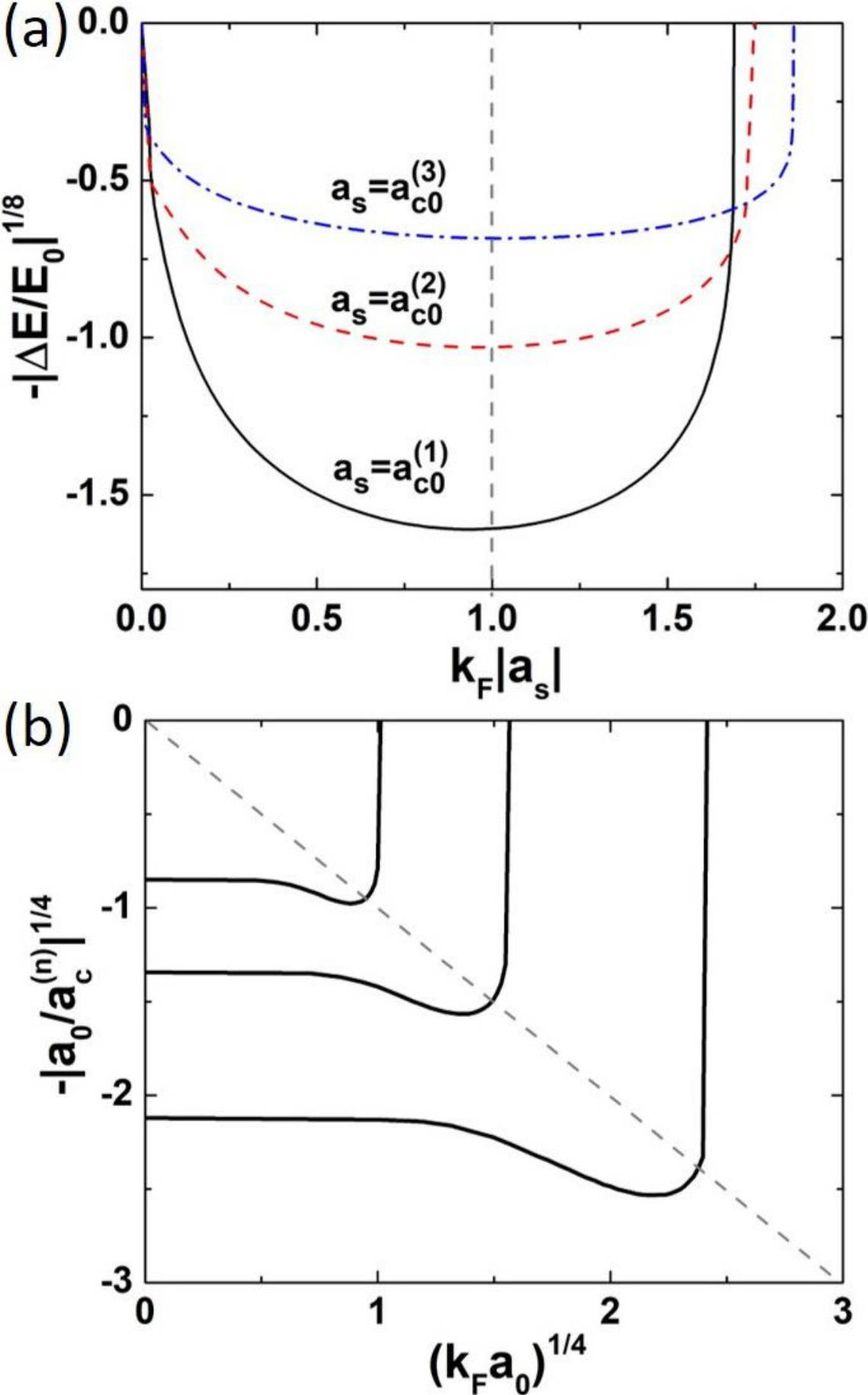}
\caption{(Color Online). (a) Bound state energies at three-body Efimov resonance ($a_s=a_{c0}^{(n=1,2,3)}$) as a function of $k_F$ (n=1(black solid line), 2(red dashed line) and 3 (blue dash-dotted line)). 
(b) Critical coupling strength $1/a_c^{(n)}$ (i.e., bound state threshold)  as a function of $k_F$. Gray dashed lines in (a) and (b) mark the location of $k_F |a_s| =1$ for the eye guide.}    \label{fig2}
\end{figure}

The above feature can be seen more clearly in Fig.\ref{fig2} (a,b), where the variations of $\Delta E$ at the three-body Efimov resonance (i.e., $a_s=a_{c0}^{(n)}$) and the critical $1/a_c^{(n)}$ are plotted as a function of $k_F$. We can see that  both $\Delta E$ and $1/a_c^{(n)}$ evolves non-monotonically as increasing $k_F$. Accordingly, there are two different regimes. In the regime $k_F|a_s|\lesssim1$, which refers to small Fermi sea and weak boson-fermion coupling, the Fermi sea facilitates the bound state formation in that $|\Delta E|$ gets larger and $1/a_c^{(n)}$ moves further to the BCS (Bardeen-Cooper-Schrieffer) side (i.e., $1/a_s \rightarrow -\infty$) as increasing $k_F$. While in the regime $k_F|a_s|\gtrsim1$, which refers to a large Fermi sea and strong boson-fermion coupling, the Fermi sea disfavors the bound state formation, in that $|\Delta E|$ get reduced and $1/a_c^{(n)}$ moves to resonance regime as increasing $k_F$. These properties of $\Delta E$ and $a_c^{(n)}$ reveal distinct effects of the fermion background to the formation of Efimov-type bound states, depending on the dimensionless parameter $k_F|a_s|$. 

\begin{figure}[t] 
\includegraphics[width=8.0cm,height=9.0cm]{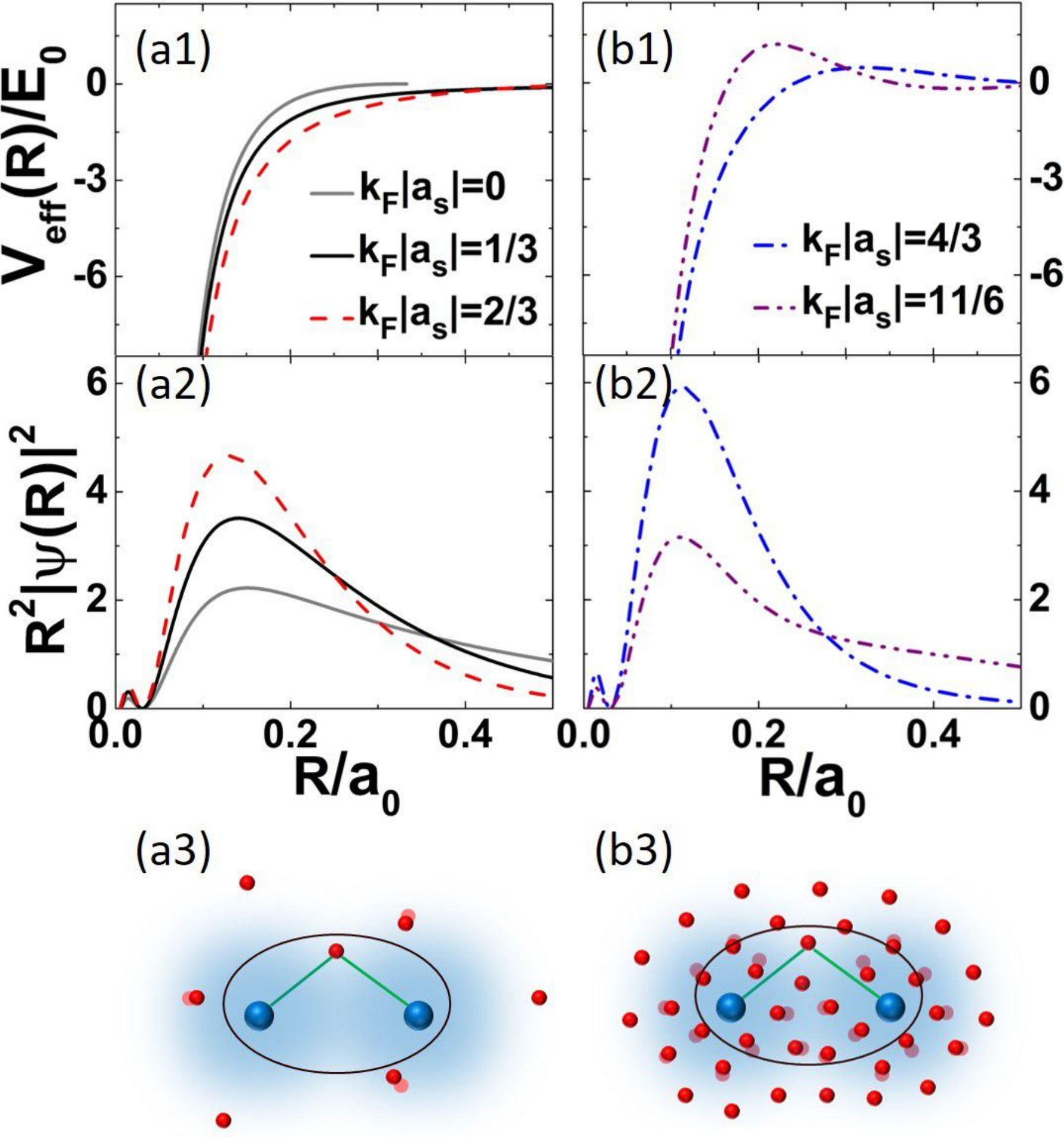}
\caption{(Color Online). The fermion-mediated interaction $V_{\rm eff}(R)$ (a1,b1) and   the bound state wave function $\psi(R)$ (a2,b2) as a function of relative distance $R$ between two heavy bosons. Here we consider a shallow bound state at the same $a_s=-a_0/3$ for all figures;  (a3,b3) show corresponding schematic pictures for the size of three-body bound state $\sim |a_s|$ in comparison with the inter-fermion distance $l$. (a1,a2,a3) are for small fermion densities in the regime $k_F|a_s|\lesssim1$: $k_F|a_s|$ =1/3 (black solid line) and 2/3 (red dashed line); while (b1,b2,b3) are for large fermion densities in the regime $k_F|a_s|\gtrsim1$: $k_F|a_s|$=4/3 (blue dash-dotted line) and 11/6 (purple dash-dot-dotted line). In addition, we show the case of zero $k_F$ in all figures as a reference (gray lines). 
} \label{fig3}
\end{figure}

To explain these distinct features, we examine the fermion-mediated interaction $V_{\rm eff}(R)$ and the resulted bound state wave function $\psi(R)$ at different fermion densities (or different $k_F$), as shown in Fig.\ref{fig3}. Here we have taken a shallow bound state at a given negative $a_s=-a_0/3$. In the three-body case ($k_F=0$), $V_{\rm eff}$ is purely given by $-\kappa^2/(2m)$, which is non-zero only in the regime $R<|a_s|$ and thus the wave function weight $R^2|\psi(R)|^2$ dominates in this region. When turning on a small $k_F$, as seen from Fig.\ref{fig3}(a1,a2), $V_{\rm eff}$ will be further lowered in the region $R<|a_s|$, which leads to a more pronounced $\psi(R)$ and thus a deeper bound state with larger $|\Delta E|$. However, when $k_F$ is increased to be fairly large, as shown in Fig.\ref{fig3}(b1,b2), the Fermi sea plays a completely different role. In this case, $V_{\rm eff}$ starts to become positive at $R\gtrsim 1/k_F$, and keeps oscillating at long distance\cite{footnote2}. As increasing $k_F$, the region of repulsive $V_{\rm eff}$ becomes more pronounced at distance  $R\lesssim|a_s|$, and accordingly the ground state wave function $\psi(R)$ gradually become broadened and extend to long distance. In this case, the bound state will be disfavored by the Fermi sea and $|\Delta E|$ get reduced. These  behaviors in $V_{eff}$ and $\psi$ explains the non-monotonic $\Delta E$ and $1/a_c^{(n)}$ as functions of $k_F$, as revealed in Fig.\ref{fig2}.

From the expression of $V_{eff}(R)$ (Eqs.\ref{energy},\ref{potential}), we can see that the effect of Fermi sea all comes from the scattering part, i.e., the collective contribution from the phase shifts of scattering states.  For $k_F|a_s|\lesssim1$, the scattering states gives an overall attractive mediated interaction between bosons (see Fig.\ref{fig3}(a1)); while for $k_F|a_s|\gtrsim1$, the repulsive component will become more dominant in the mediated interaction (see Fig.\ref{fig3}(b1)). 
A physical picture to understand this is the following.  Take a shallow Efimov bound state, its typical size $l$ is roughly determined by $a_s$, i.e. $l\sim |a_s|$, since the attractive potential $V_{eff}(R)$ ends near $R\sim|a_s|$ (see Fig.\ref{fig3}(a1)). Therefore, for a dense Fermi gas with inter-particle distance smaller than the size of three-body bound state, $1/k_F<l\sim |a_s|$ (see schematic plot in Fig.\ref{fig3} (b3)), many fermions will enter the region of three-body wave function, and the scattering process will be strongly blocked by the dense fermions, which disfavors the bound state formation. In this case, our result is consistent with previous studies showing the Pauli-blocking effect can indeed suppress the bound state formation in similar setup\cite{Zhou, Zinner1}. This situation is in contrast to another limit for a dilute Fermi gas, where $1/k_F>l\sim |a_s|$ (see schematic plot in Fig.\ref{fig3} (a3)) and the Pauli-blocking effect is negligible. In this case,  the fermions contribute an overall attractive interaction between bosons. 
Namely, in the regime $|a_s|\ll 1/k_F,R$, and up to the order of $a_s^2$, $V_{\rm eff}(R)$ follows the type of RKKY(Ruderman-Kittel-Kasuya-Yosida) potential\cite{Nishida1}
\begin{equation}
  V_{\rm eff}(R)=\frac{a_s^2}{4\pi m}\frac{2k_F R\ {\rm cos}(2k_F R)-{\rm sin}(2k_F R)}{R^4}
\end{equation}
By integrating it in coordinate space, we obtain an overall attractive interaction $\bar{V}_{\rm eff}=-k_F a_s^2/m$. This is consistent with the negative effective boson-boson scattering length in the weak coupling limit\cite{Chin,finduced_th1,finduced_th2,finduced_th3}.

\begin{figure}[t]  
    \includegraphics[width=8.0cm,height=6.50cm]{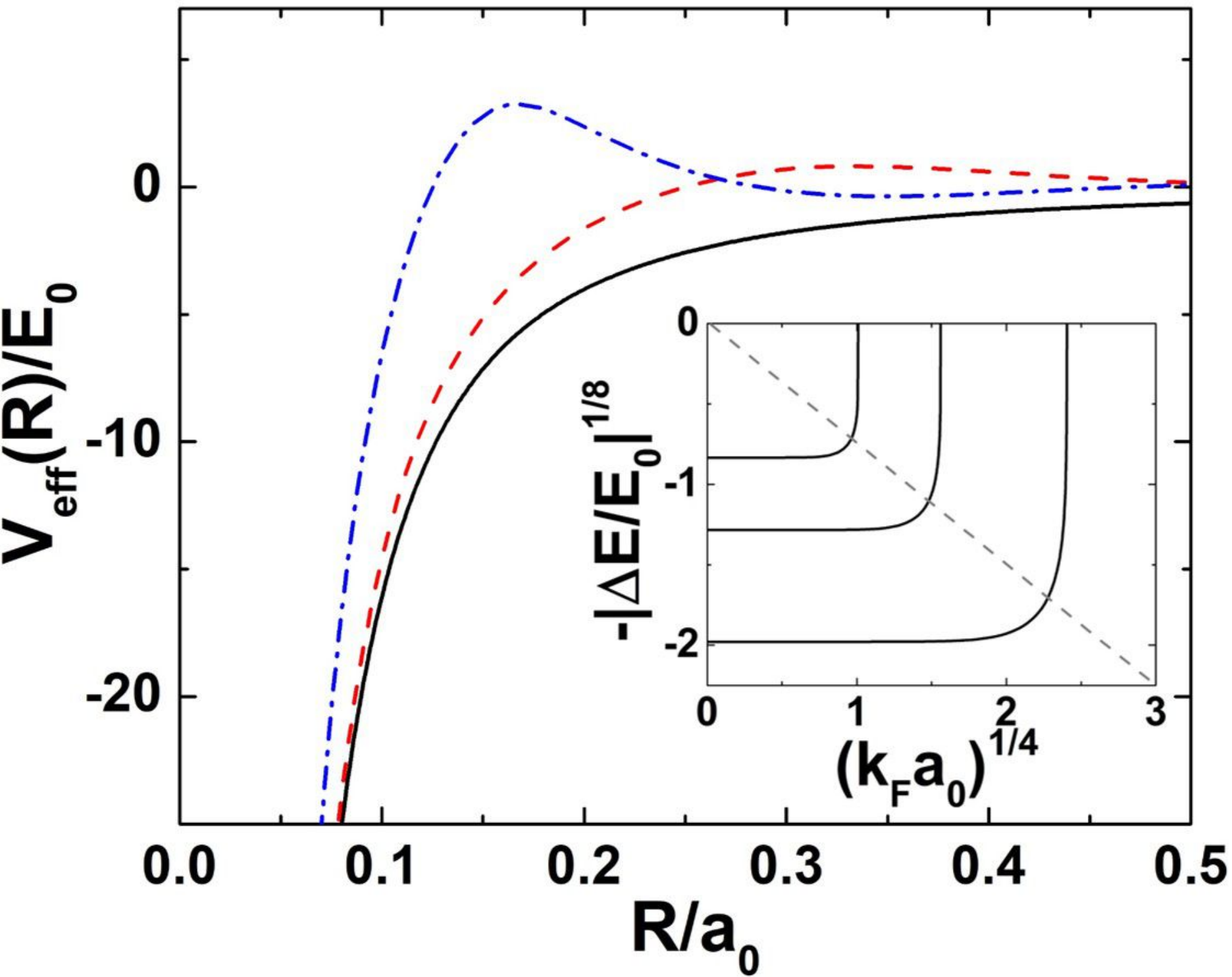}
    \caption{(Color Online). Effective potential $V_{eff}(R)$ mediated by fermions. The Fermi momentum $k_F a_0$=0(black solid line), 3(red dashed line) and 6(blue dash-dotted line) respectively. Inset: the bound state energy $\Delta E$ as varying $k_F$ at resonance $1/a_s=0$. The gray dashed line is for the eye guide. 
    }
     \label{fig4}
\end{figure}

Near unitarity ($k_F|a_s|\gg 1$), above picture immediately predicts that the bound state formation is essentially disfavored by the Pauli-blocking effect of fermions. In Fig.\ref{fig4}, we show $V_{eff}(R)$ and the resulted bound state spectrum for resonant boson-fermion interaction. Indeed, we can see that as increasing $k_F$, $V_{\rm eff}$ is shifted upward (Fig.\ref{fig4}), and accordingly $|\Delta E|$ decreases and finally vanishes at certain $k_F$ (Fig.\ref{fig4}(inset)).

A remarkable feature of the Efimov physics is its universal discrete scaling law. Here, in the presence of an underlying Fermi sea, a new momentum scale $k_F$ enters the problem and the original scaling law breaks down. However, one can establish a new set of scaling laws by incorporating the variation of $k_F$. Previously, part of the scaling laws were unveiled using different techniques\cite{Zhou,Zinner1}. Here we present the generalized discrete scaling relations for both the binding energies and the appearance locations of the dressed bound states. In the framework of BOA, this is seen more clearly in the unitary limit, where $\delta_0(k)=\pi/2$ and ${\rm tan}\delta_{\pm}=\mp(k'k_Fr\pm {\rm tan}(k'k_Fr))/{\rm cos}(k'k_Fr)$ with $k'=k/k_F$, the induced interaction $V_{\rm eff}$ falls into the following form:
\begin{equation}
   V_{\rm eff}(R)\propto -\frac{0.567^2}{2R^2}-k_F^2 f(k_FR),
   \label{potential2}
\end{equation}
here the dimensionless function $f(x)$ can be expressed as
\begin{equation}
\begin{aligned}
f(x)=& \frac{1}{\pi}\int_0^1 dy y \{ {\rm arctan[-\frac{xy+{\rm tan}(xy)}{{\rm cos}(xy)}]} \\
    &+{\rm arctan[\frac{xy-{\rm tan}(xy)}{{\rm cos}(xy)}]} \} -\frac{1}{2}
\end{aligned}
\end{equation}
Note that here ${\rm arctan}[x]$ is in the range [0,$\pi$].  
Then if we transform $r\rightarrow \eta r$ and $k_F\rightarrow k_F/\eta$ simultaneously, the Schrodinger equation (Eq.~\ref{SchrodingerEq}) recovers the continuous scaling symmetry. Further,  one can develop a new discrete scaling law similar to that in the Efimov physics. Namely, by doing the scaling transformation
$\mathbf{R}\rightarrow \lambda^{-1}\mathbf{R}, \  a\rightarrow \lambda^{-1}a,\ k_F\rightarrow \lambda k_F$, where $\lambda=e^{\pi/s_0}$ is the scaling factor, the wave function $\Psi$ still satisfies the Schr\"{o}dinger equation with energy $E\rightarrow \lambda^{2}E$. This results in the scaling law for the bound state energy:
\begin{equation}
\frac{E_{n+1}(k_F,a)}{E_n(\lambda k_F,\lambda^{-1}a)}= \frac{1}{\lambda^2}. \label{scaling1}
\end{equation}
Similarly, for the location of these bound states we have
\begin{equation}
\frac{a_{c}^{(n+1)}(k_F)}{a_{c}^{(n)}(\lambda k_F)}= \lambda. \label{scaling2}
\end{equation}
They can be seen as the generalized discrete scaling law in the many-body environment.We can verify these relations by comparing the x and y coordinates of the  intersections between any radial line and the curves shown in Fig.\ref{fig2}(b) and Fig.\ref{fig4}(inset). For example, in Fig.\ref{fig2}(b), the locations of three crossing points between gray line and solid lines satisfy the geometric scaling relation (\ref{scaling2}).   

\textit{Summary and discussion.} In summary, we have studied the Efimov correlation in a system of two heavy bosons immersed in a bath of light fermions. Taking the concrete model of Li-Cs mixtures, our results reveal the non-trivial many-body effect to the formation of Efimov-type bound states. The non-monotonic behavior of the bound states as increasing the fermion density,  as well as the generalized discrete scaling law in the fermion density domain,  can all be realistically detected by measuring the modified bound state spectrum\cite{rf_1,rf_2} and the shifted Efimov resonance\cite{Efimov_Exp0,Efimov_Exp1,Efimov_Exp1bu,Efimov_Exp3, Efimov_Exp5,Efimov_Exp9,Efimov_Exp10,scaling_1,scaling_2,scaling_3} as done previously in cold atoms experiments in the few-body context. For instance, for the $^6$Li-$^{133}$Cs mixture at the second three-atom threshold $a_{c0}^{(2)}\sim -100$nm\cite{scaling_2, scaling_3}, when increasing the fermion density to $\sim 2\times 10^{13}\ {\rm cm^{-3}}$, the new Efimov resonance will move to $a_c^{(2)}\sim -50$nm.

It is interesting to ask whether the non-monotonic effect of Fermi sea is still robust without the BOA used in this work, or, whether it can be extended to systems with a general mass ratio between bosons and fermions. Our preliminary answer is yes. This is based on two observations. First, in the weak coupling regime (small $|a_s|$), the effective boson-boson interaction mediated by fermions is always attractive\cite{finduced_th1,finduced_th2,finduced_th3}, regardless of the mass ratio.  Secondly, in the case of very dense fermions (large $k_F$), it is natural from the schematic picture (see Fig.\ref{fig3}(b3)) that the bound state will be strongly affected (disfavored) by Pauli-blocking effect. Given that $k_F|a_s|$ is the only dimensionless long-range parameter in the problem, these two situations directly determine two different regimes for the Fermi sea effect on bound state formation. Namely, the limit of small (large) $k_F|a_s|$ will favor (disfavor) the formation of dressed bound states, which holds true even outside the validity regime of BOA to any mass ratio between impurity bosons and background fermions.

{\it Acknowledgements.} We thank Xin Chen for helpful discussions. The work is supported by the National Key Research and Development Program of China (2018YFA0307600, 2016YFA0300603), and the National Natural Science Foundation of China (No.11622436, No.11421092, No.11534014).  


\begin{thebibliography}{99}
\bibitem{footnote} The experiment in Ref.~\cite{Arhus} reveals the three-body (not Efimov) contribution in the polaron spectrum.
\bibitem{footnote1} The lowest branch bound states are hardly affected by the Fermi sea as their sizes ($\lesssim |a_s^{(1)}|$) are much smaller than the inter-fermion distance ($\sim 1/k_F$).
\bibitem{footnote2} Note that the oscillating potential $V_{\rm eff}$ in the regime $k_F|a_s|\gtrsim1$ should be distinguished from the RKKY potential, which applies in the region $|a_s|<<1/k_F, \ R$\cite{Nishida1}.
\end{thebibliography}
\end{document}